
\input epsf
\input harvmac.tex
\noblackbox

%
%
\ifx\epsfbox\UnDeFiNeD\message{(NO epsf.tex, FIGURES WILL BE IGNORED)}
\def\figin#1{\vskip2in}
\else\message{(FIGURES WILL BE INCLUDED)}\def\figin#1{#1}
\fi
\def\Fig#1{Fig.~\the\figno\xdef#1{Fig.~\the\figno}\global\advance\figno
 by1}
%
%
%
%
\def\ifig#1#2#3#4{
\goodbreak\midinsert
\figin{\centerline{\epsfysize=#4truein\epsfbox{#3}}}
\narrower\narrower\noindent{\footnotefont
{\bf #1:}  #2\par}
\endinsert
}
\lref\PeeblesA{P.J.E.P. Peebles Physical Cosmology, 1980}
\lref\NTA{N. Turok, Phys. Rev. Lett. {\bf 63} (1989) 2625 }
\lref\NTB{N. Turok, Princeton preprint PUP-TH-1230
 to appear in ''The Birth and Evolution of Our Universe'',
 proceedings of Nobel Symposium, Sweden }
\lref\Kibble{Kibble T.W.B., J. Phys. {\bf A9} (1976) 1387 }
\lref\STPR{D.N. Spergel,N. Turok,  W.H. Press, B.S. Ryden , Phys Rev D}
\lref\Tursper{N.Turok and D.Spergel, Phys. Rev. Lett. {\bf 64} (1990) 2736 }
\lref\Good{A. K. Gooding, ApJ Letters, (1991) }
\lref\Goodth{A. K. Gooding, Princeton University PhD thesis (1991) }
\lref\Cen{R.Y. Cen, J.P. Ostriker, D.N. Spergel, N. Turok, ApJ. (1991) }
\lref\OrigA{J.S., Hall Science {\bf 109} (1949) 166 }
\lref\OrigB{W.A., Hiltner Science {\bf 109} (1949) 165 }
\lref\Ruzmaikin{I.N. Mishustin and A.A. Ruzmaikin , JETP {\bf 34} (1972) 233 }
\lref\Ryden{B. S. Ryden PhD thesis , Princeton 1987 }
\lref\Parker{E.N. Parker,
Cosmical Magnetic Fields, 1979 Clarendon Press- Oxford}
\lref\Moffatt{H.K. Moffatt
Magnetic Field generation in electrically conducting fluids
( Cambridge U. Press, 1978)}
\lref\Krause{F. Krause and K.-H. R\"{a}dler,
Mean Field MHD and Dynamo Theory,1980 Pergamon Press}
\lref\Beck{Ed. R. Beck, , P.P. Kronberg and R. Wielebinski ,
Galactic and Intergalactic Magnetic Fields,
IAU Symposium no. 140, 1989 Kluwer Academic Publishing}
\lref\ParkerB{E.N. Parker , ApJ {\bf 401} (1992) 137}
\lref\KulsrudAnderson{R. M. Kulsrud and S. W. Anderson ,
ApJ {\bf 396} (1992) 606}
\lref\KulsrudB{R.M. Kuldrud, IAU Symposium no. 140,
1989 Kluwer Academic Publishing}
\lref\Dynamo{A.A. Ruzmaikin  ''Galactic and Intergalactic Magnetic fields'',
Kluwer Academic Publisher (1990) p. 88 }
\lref\HarrisonA{E.R. Harrison, Mon. Not. R. astr. Soc. {\bf 147} (1970) 279 }
\lref\HarrisonB{E.R. HArrison, MNRAS {\bf 165} (1973) 185}
\lref\ratra{B. Ratra, Ap. J {\bf 391} (1992) L1}
\lref\Daly{Daly, R, Loeb, A. Ap. J. {\bf 364}(1990) 951}
\lref\derrick{Derrick 1964}
\lref\hobard{Hobard 1963}
\lref\Dolgov{A.D. Dolgov and Sun Hong Rhie, preprint CfPa-TH-92-015 (1992)}
\lref\Vachaspati{T. Vachaspati, Physical Review D, {\bf 45} (1992) 45}
\lref\UeLiPen{Ue-Li Pen,D.N. Spergel and N.Turok, PUP-TH-1375 (1993)}
\lref\RuzmaikinB{A.Ruzmaikin, D. Sokoloff, A. Shukurov MNRAS {\bf 241} (1989)
 1}
\lref\Browne{P.F. Browne , Astron. Astrophys. {\bf 144} (1985) 298}

\Title{PUPT-xxxx}
{\vbox{
   \centerline{Angular Momentum and Large-Scale Magnetic}
   \vskip .1in \centerline{Field in Textures Seeded Models}
   \vskip .1in \centerline{\ }}}
    \vskip .1in \centerline{Hugues Sicotte\footnote{$^\dagger$}
{work supported in part by NSF}  }
\bigskip\centerline{\it Joseph Henry Laboratories}
\centerline{\it Princeton University}
\centerline{\it Princeton, New Jersey 08544}
\centerline {\it sicotte@pupgg.princeton.edu}
\vskip .3in
\centerline{\bf Abstract}
We start by reviewing the problem of large scale magnetic fields.
We explain the acquisition of angular momentum in the general
context of Dark Matter models and we use this result to
derive a dynamical mechanism for magnetic field generation.
This mechanism does not produce any magnetic field
for standard CDM, but it does for a
large class  of other models.
We apply this mechanism in the context of the
texture scenario of large-scale structure
formation. Resulting constraints on the texture scenario and other
models are discussed.
%
%

\Date{November 1994} 
\newsec{Introduction}

\par
There is strong observational evidence for a large scale galactic
magnetic field in our galaxy and in neighbouring galaxies . From the
early polarisation of starlight measurements \OrigA \OrigB \ to the more
recent observations of the faraday rotation of pulsar emissions or other
extra-galactic sources, there is a strong experimental evidence of such
field. The galactic magnetic field is also required to explain why
cosmic ray composition makes it appear that they have been through
30 to 300 time as much matter as there is between us and the galactic
center. Alfv\'{e}n
and Fermi (1949) first discussed this and predicted
a field of $ 10 {\mu}$gauss to confine the cosmic rays in the galactic
plane, this is very close to the current $2-3 {\mu}$gauss
value for the coherent component. Current theories of this magnetic
field creation require a seed field which is then amplified by a dynamo
mechanism. \Dynamo

\par
In addition, more recent observations show that there seem to be a magnetic
field on cluster size.
Any theory of large scale structure should predict some seed magnetic
field uniform on a galactic scale, the texture mechanism \NTA does this in a
very natural fashion.

\par
The outline of this paper is as follow.
In section 2, we will describe how one generically
generates magnetic field with a model of cosmological
perturbations. In section 3, we will briefly
describe the model of large scale structure formation by textures, and
quote some of the results we will be using. In section 4 we will
apply the formalism of section 2 to the
texture model of structure formation and calculate the magnetic field
it generates. In section 5 we will
discuss the results and the implications for texture and compare to
other candidate model of structure formation.
\subsec{Notation}
When discussing physics happening around matter-radiation equality,
it is natural to refer scales to that era, so we will use a scale
factor normalised at matter-radiation equality:
$a(t), {\rm where }\/ a(t_{eq}) \equiv 1$ . Sometimes we will
 use the more common scale
factor defined as ${R \over R_0}\equiv {1 \over (1+z)}$.
Spatial distances denoted by $x$
, will always be in co-moving
coordinates in units of Mpc. $x_{eq}$ denotes co-moving
coordinates with a value that
they would have at Matter-Radiation equality,
$x$ without the eq subscript denotes
the normal co-moving coordinates, referenced to the present time.
$k$ will denote ${2 \pi \over \lambda}$ and will be in co-moving units
of $Mpc^{-1}$; here again $k_{eq}$ will be referenced to equality, and
$k$ without a subscript will be referenced to the present.
\par
Time for a co-moving observer ,$t$, will be given in units of $
ls$ . Where 1 ls$=2\sqrt{2}c\tau_{*} \approx 2.79 \times 10^{12}$ seconds
We will convert between $a$ and $t$ using
\eqn\Ta{t = {t_{eq}\over 2- \sqrt{2} }\bigl( (1+a)^{3 \over 2} -3 \sqrt{1+a}
      +2 \bigr)  }
\par
and it's numerical inverse. Variables with a subscript
i ( i.e. $a_i$ , ${\tau }_i$ ) refer to the
value of that variable at the epoch when a given texture just collapsed.
\subsec{CDM Model}
In all following discussions standard CDM will refer to a
Cosmology with a scale invariant Harrison-Zeldovich (HZ) spectrum.
For both CDM and the
Texture model we will use a model with critical density $\Omega =1$ ,
with a Dark Matter component $\Omega\sim 0.95$ and a baryonic
fraction of 5\% . For the standard CDM model, we will assume a Hubble
constant of $50 km/s/Mpc$, while for the textures model we will
assume a Hubble constant of $75 km/s/Mpc$ .
For our purposes, the exact composition of the model makes little difference as
long as most of the matter is Dark Matter.
\par
Throughout this article we will use DM when speaking
about the dark matter, while baryonic and leptonic matter and photons will be
refered to as matter.

\newsec{Magnetic Field Generation}
\par
When thinking about magnetic field generation, it is useful to separate
the physics in 2 phases.
First a seed field, $B_{seed}$, is created before a galaxy collapses. In the
second phase the magnetic field is first amplified
by flux conservation during collapse and then by a dynamo process.
\par

The seed field creation ceases after the protons and electrons
present
in the primordial plasma recombine.
%
%
%
%
%
%
The flux lines are then frozen in the interstellar gas due to
the large conductivity of the plasma.

For an introduction to the dynamo theory see \Moffatt \Krause \Parker \Beck .
The dynamo amplification mechanism occurs in galactic objects.
The flux is amplified through processes that makes
flux lines buckle and form loops
arcing above the galactic plane . These loops twist $180^\circ$,
rejoin and fall back on the galactic plane, increasing the flux line density.
 For example, a cloud
of gas can rise due to buoyancy forces, it then  expands due to the lower
density of gas above the galactic plane. As it expands, the coriolis force
makes the cloud rotate, twisting the lines of flux
as they are frozen in the cloud. The cloud then cools and falls back
onto the galactic plane, carrying the loop back in the
plane. This process increases the random field,  $B_{rms}$. The
ensemble average of these random fields has a growing coherent mode.
It is a topic of some controversy as to whether
this yields an exponential or a linear increase of the magnetic field
\KulsrudB \KulsrudAnderson \ParkerB.

To get the current value of $3 \mu$gauss,with exponential dynamo ,
 seed field values of only
about $10^{-18}$  to $10^{-16} $gauss are required at the time
of galaxy formation.
 On the other hand a linear dynamo requires fields of at least
$10^{-12} \mu$gauss which are very hard to generate without delicate
adjustement of large-scale structure generating models. For such an example
in the inflationary scenario see \ratra
\par

The creation of a seed field of even $ 10^{-18}$ gauss is a non-trivial
test for a theory to pass. Generically seed field creation can
happen in 4 possibles eras. The first one is if the magnetic
field has a primordial origin, for example \ratra .
The next two other eras involve dynamical phenomenas. The last
period happens once the structure formation is well advanced. It
involves magnetic field expulsed from powerful beamed radio
sources.\Daly
\par
We will not deal here with theories postulating a primordial of
the magnetic field or those postulating the existence of
powerful radio sources at very high redshift.

\subsec{Harrison scenario}
The second era for magnetic field generation
is between matter-radiation equality
and recombination. If we assume that there was some primordial
vorticity, such as could happen in chaotic Cosmology,
we can generate magnetic fields using to the following mechanism
due to Harrison \HarrisonA \HarrisonB .

During that era the electrons motion is tied to that
of the photons though the large thompson cross-section,
$\sigma_e^T$.
For the protons ,the proton's thompson cross section,
$\sigma_p^T$, is
much less important than the hadronique neutron-proton cross
section. As a result while the electrons move with the photon
fluid, the protons together with the neutrons can have a different
behavior.
\par
Implicit in the derivation of the Harrison mechanism is the
existence of primordial vortices.
We can nevertheless define a local scale factor $r$
within the framework of the Homogeneous world model. The eddy follows
the expansion and
$\rho r^3$ and $\rho r^4$ remain constant for matter and radiation
respectively. So the angular momenta of the matter, $\rho \omega
r^5 $, and
the angular momenta of the radiation, $\rho_r \omega_r r^5 $, are separately
conserved. Therefore $\omega_p \propto {1 \over r^2}$ and $\omega_{\gamma}
 \sim \omega_e \propto
{1 \over r}$ . This means that electrons and protons will have different
angular velocities, leading to an angular electric field and therefore
a polo\"{\i}dal magnetic field. The value of this field will
get frozen in at recombination.

The detailled kinematical analysis can be found in
Appendix 1. The final result is
\eqn\BHarrison{B_{recombination} = {-2m_p \over e}{\bf w}(z_r)=
(-7.46 \/ 10^{-17} gauss (bs)) {\bf w}(z_r) (bs)^{-1}}

In an expanding Universe, the initial vorticity
($\zeta = \nabla \times \vec{v}$ ) perturbations decay as

\eqn\zetaexpand{\zeta (a) \sim  {1 \over a^2} }.

This mean initial vorticity is washed away unless the universe was
very chaotic when it started. Limits on the amplitude of
primordial density perturbations by the COBE satellite
rule out this possibility unless the universe was re-ionized.

\subsec{Harrison-like scenario}
The above Harrison scenario fails to pass the experimental tests
because it depends on primordial vorticity. We will ressucite
the Harrison mechanism, but rely on dynamically generated
vorticity. This will yield the third period when magnetic field
can be generated.
The vorticity in the matter will be generated when the DM
collapses early and acquires non-spherical components through
tidal torquing. The acoustic waves in the matter,
 that form as the super-horizon
density modes enter
the horizon, will then undergo oblique shocks because they are
falling in a non-spherical rotating potential. This mechanism
will thus create the required vorticity in the matter.

In most theories of Large Scale Structure (LSS) formation
the Jeans mass before decoupling is larger than the mass within
the Horizon. This is due to the photon pressure which
keeps density perturbations initially at rest  from growing .

But LSS theories with topological defects
 generate density perturbations and velocity until very late
times,
 unlike standard LSS  scenarios
which have all their density perturbation as initial conditions.

In the texture model the density perturbations are not fixed as
initial condition. The density perturbation are dynamically
generated because of the texture collapse.
We will see later that the regions around the center of
a collapsing texture get a velocity kick toward that center.
The radial component can be approximated by

\eqn\veltexture{{\rm \vec{v} }= \epsilon c
\bigl(1- {{\rm r} \over c t_i} \bigr) }
where r is the real radius, $t_i$ is the collapse time of the texture and
$\epsilon $ is the only parameter of the theory and is normalised
to COBE. This radial kick allows a texture seeded velocity kick to
grow density even though the matter contained within it is lower
than the Jeans mass before decoupling.

The velocity kick  is calculated for an ideal
spherically symmetric texture collapse.
Numerical textures simulations show that most textures are not
perfectly symmetric and that a small azimuthal velocity is also imparted
to the matter surrounding the texture. Typically there are non-radial
components to the velocity of approximatively 1 -2 \% of \veltexture .
This is not to say that a vorticity is created, it is a
perturbation from a purely radial velocity kick due to the
assymetric nature of real textures.
\par
We will not be
using the small rms azimutal velocity from the initial
collapse for our final result.
Instead the initial assymetries in the dark matter, inside the
volume of the texture, will be caused by the previous
collapse of the smaller textures that were contained in that
volume. We will see in section 4 how those assymetries
cause the DM halo to gain angular momentum through tidal torques
from neighbouring DM halos.
We will see how this rotating DM halo
will then transfer angular momentum to the
matter before the recombination era.
This allows some magnetic field generation for late textures
by a mechanism similar to the Harrison scenario.

\par

\newsec{Large-scale structure texture seeds}
\par
The theory of topological defects as source of large scale structure was
first treated by Kibble \Kibble , later
Turok \NTA showed that global texture could
occur naturally in GUT theories and produce interesting density
perturbations in the early universe.
(See \NTB\Goodth for review)

\par
An SU(2) symmetry seems most natural in term of a particle
physics model. With a complex
higgs field we could have a global
SU(2) symmetry . For  computational purposes it is easier to deal
with a SO(4) 4-component isovector with real fields, so let us
consider a GUT theory with a global SO(4) symmetry .

\eqn\lagrangian{ {\cal L} = {1 \over 2} \bigl( \partial_{\mu}\vec{\Phi}
\bigr) \bigl(\partial^{\mu}\vec{\Phi}
 \bigr) \/  - {\lambda \over 4}(\vec{\Phi}^2 - \Phi_0^2)^2  }

with \eqn\PHI{\vec{\Phi} = {1\over \sqrt{2}} \pmatrix{\Phi_1 \cr \Phi_2 \cr
\Phi_3 \cr \Phi_4 \cr } }
where the lagrangian is invariant under the global symmetry transformation
$\vec{\Phi} \rightarrow
\exp{\imath\alpha} \vec{\Phi}$ with $\alpha$ a constant.
In the ground state , $\vec{\Phi}$ develops an expectation value,
${<0\vert\vec{\Phi}{\vert}0>} = (\Phi_0,0,0,0)$,
which  breaks the symmetry of the lagrangian.
 . The 1-direction is arbitrary so that, early in the universe,
fields not in causal contact will point in different directions.
As their event horizon grows and they come in causal contact
they may form a field configuration known
as texture. This large size field configuration is unstable by derrick's
 theorem. \derrick\hobard \ which states that no static soliton solution exists
in 3 or more dimensions. So the energy stored  in the gradients of the field
configuration collapses to the center. At that point the energy
density is such as to allow the texture field configuration
to topogically unwind over the potential. This leaves the field aligned
within the current horizon volume.
As the horizon grows, other texture field configurations may
appear and collapse. Each time a texture collapses, the energy
of the field configuration produces a gravitational attraction toward
the center of the texture. The resulting radial
velocity kick can be described as

\eqn\velkick{ {\rm \vec{v} } = \epsilon c \bigl( 1- {{\rm r} \over c t_i}
\bigr) }
where $\epsilon=8\pi G \Phi_\circ^2$ is normalised to COBE.
There is a  resulting azimutal velocity kick of at most  1-2\% of
this. The resulting velocity field does not have a net vorticity.
This aximuthal component is simply due to the field configuration
not being uniformly spherical.
\par
\subsec{Power spectrum}
In \ \UeLiPen \ Pen et al.. calculated numerically the power spectrum
caused by texture as it would look today. They give a fitting function
\eqn\Powerfit{{V  P_{fit}(k) \over (2 \pi)^3} =
{890 (Mpc^{4}) k \over (1 +14.\times k +
 11\times k^{1.5})^2}}

\par
We can use the fitted function of the numerical results in
\UeLiPen and evolve it backwards in time using linear theory and
an time dependant cut-off for small $k$
\eqn\Powertur{VP(k_{comov},a) = VP(k_{comoving})
\times {{D(a)^2}\over D(a_{now})^2} }
where $k_{comov} <  {2\pi a \over a_{now} ct}$
and $D(a)=(1+ {3 a \over 2})$ is the growing mode solution
of a density perturbation valid in both the matter and radiation era.
\PeeblesA .

\newsec{Magnetic field by texture}
As we saw in section 2, we have to separate the problem of magnetic
field generation in 2 phases. In the
first phase the texture provided the initial aspherical density
perturbation. In the second phase
angular momemtum will be provided by torquing by gravitational forces.
Note that this process can be also applied to standard CDM.

\subsec{Angular Momentum}
Once the Texture collapse has seeded a velocity kick in the DM
particles, our mechanism still require that the resulting
particle distribution acquires a non-spherical distribution.
This rotation of the non-symmetric gravitational potential
will be acquired through the tidal torquing picture.
While in standard CDM the Dark matter cannot collapse this early,
the tidal torquing picture of Peebles \PeeblesA can still be
applied.
In what follows we will closely follow \Ryden who computed the
tidal torquing for
a standard CDM model with gausssian spectrum.
\par
Consider a perturbation which is growing, we can look at the evolution
of shells. A given shell will contain a fixed amount of matter and shells
will not cross at least until they are collapsing faster than the Hubble
expansion. The gravitational torque induced on a shell,
by all the surrounding matter
is.
\eqn\torquea{ \tau (x) = -G {\int_{x \in shell}
\rho ({\vec s} ) {\vec s} \times
 {\vec {\nabla}} \Phi (\vec s) d^3s} }
where
\par
\eqn\rhoeps{ \rho ({\vec s}) =
\rho_b (1+ {\delta }(s))(1+{\epsilon }({\vec s}) ) }
\par
All the non--sphericity
of $\rho $ is in ${\epsilon }({\vec s}) $ for $\vec s $
much larger than the size of the texture (i.e. the other earlier textures
that collapsed within this volume provide the
asphericity ).
\par
$\Phi$ is the gravitational potential, which can be expanded in spherical
harmonics.
\eqn\Phiharm{ \Phi (\vec s) = {\sum_{l=0}^{\infty }{4 \pi \over 2l+1}
{\sum_{m=-l}^{m=l} a_{lm}(s) Y_{lm}(\theta , \phi ) s^l}}}
where
\eqn\alm{a_{lm}(x) = \rho_b \int_x^{\infty}{ Y_{lm}(\theta , \phi)
(1+\delta (s))\epsilon ({\vec s})}
s^{-l-1} d^3s}
plugging \Phiharm into \torquea and using the fact that
\eqn\angop{{\vec x} \times {\vec \nabla}(x^l Y_{lm}(\theta , \phi )) = ix^l
{\vec L} Y_{lm}(\theta ,\phi )}
\par
where
\eqn\Lang{{\vec L}\equiv - i ({\vec x}\times {\vec \nabla}) =
\hat{x} (L_{+} + L_{-}) +i  \hat{y}(L_{-} -L_{+}) \hat{z} L_z }
\par
is the angular momentum operator familiar from quantum mechanics.
we obtain for the z component
\eqn\torquez{\tau_z = -i G M_{shell} \sum_{l=0}^{\infty}{{x^l \over 2 l +1 }}
\sum_{m=-l}^{l} m a_{lm}(x) \int{\epsilon({\vec x}) Y_{lm} d\Omega}}
\par
Let us introduce the multipole moments of the shell
\eqn\qlm{q_{lm}(x) = \int_{shell}{ Y_{lm}^{*} (\theta , \phi )s^l
\rho ({\vec s}) d^3s} = {x^l \over4 \pi }M_{shell}
\int{Y_{lm}^{*} (\theta , \phi) \epsilon ({\vec x}) d\Omega } }
where the final form is reached after substituting \rhoeps and being left
with the non-symmetric part. We can now rewrite $\tau_z$ as
\eqn\tauz{\tau_z = -i 4\pi G \sum_{l=0}^{\infty}{{1 \over 2 l + 1}
\sum_{m=-l}^{l} m a_{lm} q_{lm}^{*}}}
\par
If we retain only the quadrupole term
($ l=2$) we can compute the rms torque on a
shell.
\eqn\rmstor{\eqalign{\langle \vert {\vec \tau } \vert \rangle &=
3\langle \vert { \tau_z } \vert \rangle \cr
&= 3 ({4\pi \over 5}G)^2 \sum_{m=-2}^{2}\sum_{n=-2}^{2}m n
\langle a_{2m}(x) a_{2n}^{*}(x) q_{2m}^{*}(x) q_{2n}(x) \rangle}}
\par
with
\eqn\fouraq{\eqalign{&\langle a_{2m}(x) a_{2n}^{*}(x) q_{2m}^{*}(x)
q_{2n}(x) \rangle = \cr
& {\rho_b^2 M^2_{shell} x^4 \over (4\pi )^2} \int_x^{\infty}{dr_1 \over r_1}
\int_x^{\infty}{dr_2 \over r_2} \int  \int \int \int
Y^{*}_{2m}(1)Y_{2m}(2) Y_{2m}(3)
Y_{2n}^{*} \langle \epsilon_1 \epsilon_2 \epsilon_3 \epsilon_4 \rangle
d\Omega_1 d\Omega_2 d\Omega_3 d\Omega_4  }}

the final form is reached under
the assumption that the 4 point function can be written as
\eqn\Eeee{\langle  {\epsilon}_1 {\epsilon }_2 {\epsilon }_3
{\epsilon }_4 \rangle =
\langle {\epsilon }_1 {\epsilon }_2 \rangle \langle {\epsilon }_3
{\epsilon }_4 \rangle \ + \
\langle {\epsilon }_1 {\epsilon }_3 \rangle \ + \
\langle {\epsilon }_2 {\epsilon }_4  \rangle \ + \
\langle {\epsilon }_1 {\epsilon }_4 \rangle \langle {\epsilon }_2
{\epsilon }_3 \rangle \rangle}

\par
which is certainly true for a random gaussian process \PeeblesA , and a
reasonnable approximation in our case.
Then after using the definition of the power spectrum
in terms of the 2 point function
\eqn\epep{\eqalign{& \int \int Y_{2m}^{*}(1) Y_{2n}(2) \langle
\epsilon_1 \epsilon_2
\rangle d\Omega_1 d\Omega_2 = \cr
& {1 \over (2\pi )^3} \int d^3k P(k) \int e^{-i{\vec k}\cdot{\vec x_1} }
Y_{2m}^{*}(1) d\Omega_1 \int e^{-i{\vec k}\cdot{\vec x_2} }
Y_{2n}^{*}(2) d\Omega_2  } }
,the expansion of a plane wave in sperical harmonics
\eqn\expsphe{e^{-i{\vec k}\cdot{\vec x} } = 4\pi \sum_{l=0}^{\infty}
i^l j_l(kx) \sum_{m=-l}^{m=l}
Y_{lm}^{*}(\theta , \phi) Y_{2m}(\theta_k , phi_k)}
and  orthogonality relations one gets
\eqn\Torque{ \tau (x) \equiv <|\tau(x)|^2>^{1 \over 2} =
\ \ \sqrt{30} \Bigl( {4 \pi G \over 5} \Bigr)
\Bigl[ <a_{2m}^2(x)><q_{2m}^2(x)
-\  <a_{2m}(x) q_{2m}(x)>^2 \Bigr]^{1 \over 2} }
\par
where

\eqn\Qm{ <q_{2m}^2(x)> = {x^4 \over (4 \pi )^2} V M_{sh}^2
\int{k^2 dk P(k, \tau ) (j_2(kx))^2 } }

\eqn\Am{ <a_{2m}^2(x)> = {2 \rho_b(\tau ) x^{-2} \over \pi } V M_{sh}^2
\int{ dk P(k, \tau ) (j_1(kx))^2 } }

\eqn\AmQm{ <a_{2m}(x)q_{2m}^{*}(x)> = {x  \rho_b(\tau ) V M_{sh}\over 2 \pi^2}
\int{k dk P(k, \tau ) j_1(kx) j_2(kx) } }

we find that the volume factor exactly cancels out
the volume factor of the power spectrum.
for the sake of brievety,
we omitted the $\tau ,\tau_i $ dependance.
Putting it back, and integrating we find

\eqn\W{ {J(x,a,a_i) \over M_{shell} x^2} = w(x,a,a_i) =
{3 t_{eq} \over 2 ( 2- \sqrt{2} )}
\int_{a_{cutoff}}^a{da' a' \tilde{\tau} (a',x_0,a_i) \over
\sqrt{1 + a'} x(a,a_i)^2 M_{shell}}  }

The $M_{shell}$ factor cancels exactly the Mass factor in \Torque.
$\tilde{\tau}$ is given by
\eqn\tildetau{{ \tilde{\tau}(x_0,a',a_i ) \over M} =
{ {\tau (x(x_0,a_i,a'),a') \over  M}}  }

and $\tau$ is given by \Torque . $x(x_0,\tau_i,a')$
is the comoving position of a shell that was at position $x_0$ when the texture
collapsed at conformal time $a_i$ .

We can therefore use the texture power spectrum \Powertur to figure out the
angular momentum produced by gravitational torquing.
We also need the position of a shell as a functions of time.
In the Texture model, Texture collapse before recombination, even
in the radiation era. Collapse is modeled in the radiation era by
\eqn\xrad{x(x_0,a,a_I) = \Bigl({\rho_b(a) \times (1+\delta(a,a_i))\over
\rho_b(a) }\Bigr)^{-1 \over 4}}
We have used the $\delta(a,a_i) $ derived by \Goodth .
When the shell enters the matter era, we match to the
spherical collapse model \PeeblesA by matching the peculiar velocity
of the collapsed shell.
Our integrals for the torque then runs until the turn-around time, as
given by the spherical collapse model.

The angular momentum thus calculated
is not directly equivalent to vorticity. It can generate a solid body-like
rotation of the dark matter fluid sphero\"{\i}d but it has no net vorticity.
To convert
this rotation into vorticity, the fluid should undergo dissipative processes
or shocks. While the DM cannot undergo dissipative processes, it
is still possible to convert the angular momentum from the DM
shell into vorticity for the matter.
\par
As new modes of the density fluctuations enter the horizon, the
matter mode grows until it is pressure supported, and undergoes a
few cycles of dissipatives acoustic oscillations. \PeeblesA It is those shock
waves, oscillating in the rotating aspherical gravitational
potential that transfer the angular momentum to the matter.
Obtaining the efficiency of this highly non-linear phenomena at
transfering vorticity
will await our future developpement of a non-linear hydro code to
test the details of this model. Nevertheless we will make useful
approximations.
\par

The torque is caused by a quadrupole-quadrupole interaction,
whose strenght  scales as
$\sim  a^{-6}$, and whose
amplitude, proportional to P(k), goes as
$\sim (1+{3 \over 2}a)^2$. The net result is that
the torque is much more effective at early times.

This is reflected in our results. Fig 1. shows the angular velocity
acquired by an object of mass $2\times 10^{11} M_{\odot}$ seeded
at scale factor $a_i = 0.032$ at a comoving radius of 1Mpc.
Most of the angular velocity of the DM is acquired early on.
\ifig{\Fig\wa}{angular velocity at a=50 for a texture collapsing at
$0.032a_{eq} $
 the units are ($ls^{-1}$) 1 $ls=$ 2.79E12 seconds. The comoving radius is
1.014 Mpc and this shell contains $2\times10^{11}M_{\odot}$
}{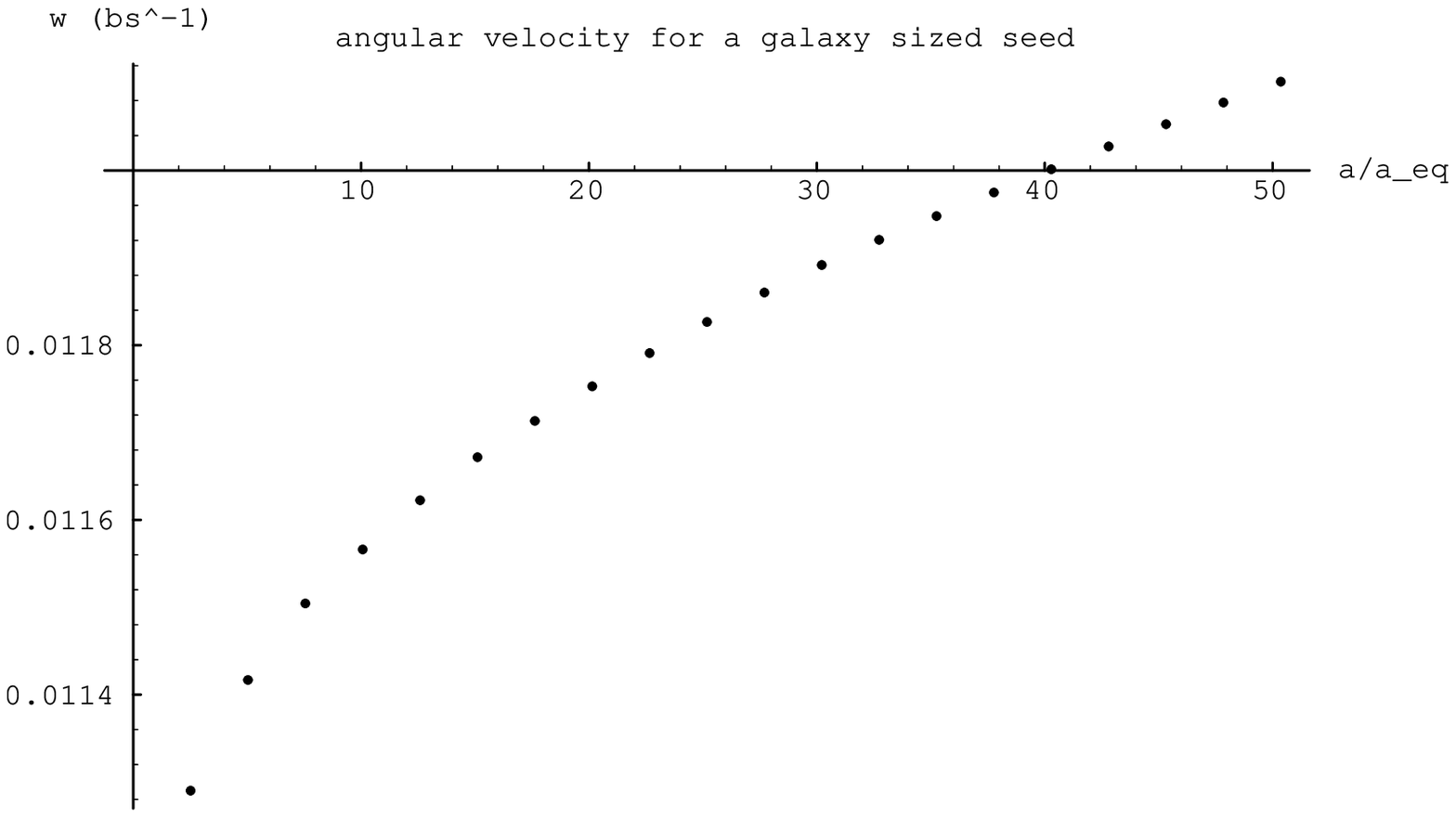}{5.5}
In Fig 2. we have plotted the angular velocity attained by each
shell at their own turn around
time. Note that the steep distribution at low radius is made steeper because
by the time the outer shells have collapsed, the center will have collapse even
further, causing a spin up if angular momentum is conserved. This could
possibly slow the collapse of the center.
\ifig{\Fig\wx}{angular velocity at the turn-around time as a function of
the unperturbed comoving radius for a texture collapsing at
$0.032a_{eq} $. Note that each $x_{ta}$ is reached at a different time.
 the units are ($ls^{-1}$) 1 $bs= 2.79\times 10^{12}$ seconds.
A comoving radius of
contains $2\times 10^{11}M_{\odot}({x \over 1Mpc})^3$
}{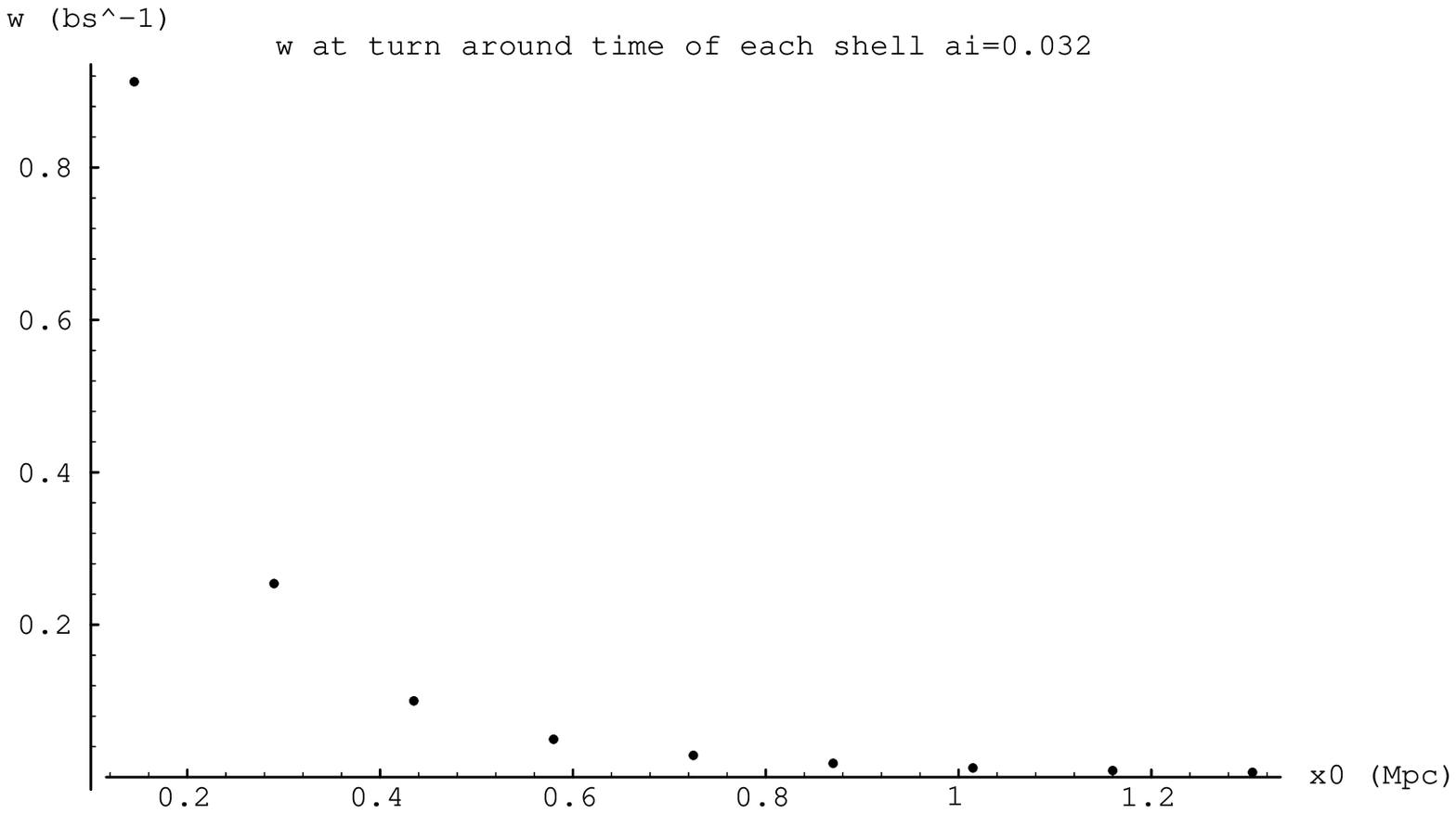}{5.5}
In table 1  we have tabulated data. The angular velocity
as a function of initial comoving radius follows
roughly $w \sim {1 \over x^{5 \over 3} }$ and
$w \sim {1 \over a_i^{8 \over 3}}$
we can compute the magnetic field for different size textures
from \BHarrison . After the radiation era, $w \sim a^{1 \over 30} $ so
we can take w approximatively constant.
As noted in the appendix we need the angular
velocity of those shells that reached their turn around point
before recombination( $a\sim 5$).
This gives that for a typical galaxy-sized object
($a_i = 0.048$,$x_0=1Mpc$,$z_f=116$, $\Rightarrow
M_{gal}=10^{12}M_\odot $ ) only regions inside a comoving radius
of 0.2 Mpc have reached
their turn-around radius. For those radii
we obtain $ \omega \sim
0.3 ls^{-1} $. To compute a magnetic field we need to know
how efficiently has the DM transfered it's angular momentum to
the matter component. Let us get an upper limit to the possible
magnetic field. Assuming 100\% conversion,  the matter acquires an angular
velocity of $w$.
We use \BHarrison to get
 $B \sim 2\times 10^{-17}$ gauss.  At that ealy epoch, the density of the
proto galaxy is actually bigger than that of the final object.
Upon complete collapse to the density
of a galaxy, this field will
be multiplied by
$ ({\rho_{collapse} \over rho_{galaxy}})^{2/3} \sim 3 \times 10^{-2}$
So textures can yield at most a seed field of $6 \times 10^{-19}$ gauss
for galaxy sized perturbations.
\par
Our galaxy has a rotational velocity of $w \sim 0.0005$ in our
units. The generated angular velocity for a galaxy-sized object
was $ \omega \sim
0.3 ls^{-1} $, at recombination, which we have to multiply by the
windup factor, $3 \times 10^{-2}$, to yield $10^{-3}$.
 This number has to be multiplied by the
efficiency factor of conversion of DM angular momentum
into matter vorticity. This shows that the required conversion
factor cannot be as high as 100\% .but has to be less about 10\% .
Under that assumption, the angular momentum of galaxy-sized
objects produced by the texture scenario is adequate.


\newsec{Conclusion}
\par
In the early universe, the texture model can produces
sizable  angular momentum.
A similar analysys using the power spectrum
of CDM has been peeformed by Ryden \Ryden .
For both models the values are consistent with our own galaxy.

In the texture model the early formation of angular momentum
yields to formation of a magnetic field. On the other hand for
the CDM model no magnetic field can
 be generated by that angular momentum because all of it was acquired
after recombination.
Our Scenario can only produce
magnetic field in Clusters if a hiearchical model of cluster formation is
adopted. This model only produced magnetic field in Galaxy sized objects and
none in the interstellar medium other than that carried by the gas ejected from
proto-objects.

\par
The upper limit we compute for the magnetic field falls short of
the modern estimates of the required seed field. Although the
texture model fares better in this problem than the standard CDM
model, it does not pass this test. Magnetic field creation
through early AGN is a better possibility in the texture model
than in CDM because the structures form earlier.

\par
The scenario we have developped shows us that
any model that yields non-baryonic growing density perturbations
by the epoch of recombination can produce magnetic fields.

\newsec{ {\bf Acknowledgements} }

I would like to thank David Spergel, Russel Kulsrud, Jim Peebles and Bharat
Ratra for helpful discussions and references.
I also want to thank to Neil Turok for suggesting to
investigate this topic.
\vfill \eject

\vbox{\moveright 0.5 in
\vbox{\offinterlineskip
\halign{\strut\vrule\quad # \quad & \vrule \hfil\quad # \hfil
& \quad \vrule \quad # \quad \vrule & \hfil\quad # \quad \hfil \vrule \cr
\noalign{\hrule}
Table 1 & & & \cr
\noalign{\hrule}
$a_i$ & $x_0$ & a/$a_{eq}$ & w ($ls^{-1}$) \cr
\noalign{\hrule}
0.003& 0.0139606& 0.134108 &20.6876 \cr
0.003& 0.0698029& 1.11764 &3.46906 \cr
0.003& 0.139606 &6 &1.25885 \cr
\noalign{\hrule}
0.004& 0.0184639 &0.18984& 15.9377 \cr
0.004& 0.0923195 &1.58212& 2.33627 \cr
0.004& 0.184639 &8 &0.773242 \cr
\noalign{\hrule}
0.006& 0.0274655& 0.31313 &10.6319 \cr
0.006& 0.137327 &2.60968 &1.28811 \cr
0.006& 0.274655 &12 &0.334348 \cr
\noalign{\hrule}
0.008& 0.0364617& 0.450027& 7.92929 \cr
0.008& 0.182309 &3.75069& 0.789084 \cr
0.008& 0.364617 &16& 0.167428\cr
\noalign{\hrule}
0.012& 0.0546771& 0.762122& 4.83116\cr
0.012& 0.273386 &6.35207& 0.33716\cr
0.012& 0.546771 &24& 0.0595082\cr
\noalign{\hrule}
0.016& 0.0728387& 1.1155 &3.26369\cr
0.016& 0.364193 &9.2977& 0.167547\cr
0.016& 0.728387 &32& 0.0288167\cr
\noalign{\hrule}
0.024& 0.109059& 1.93168& 1.76581\cr
0.024& 0.545294& 0.805078& 0.057496\cr
0.024& 1.09059 &48 & 0.0101015\cr
\noalign{\hrule}
0.032& 0.144989& 2.87665& 0.912561\cr
0.032& 0.724945& 1.19898& 0.0274236\cr
0.032& 1.44989& 64& 0.00471943\cr
\noalign{\hrule}
0.048& 0.216372& 5.12164& 0.287895\cr
0.048& 1.08186& 42.6972& 0.0102267\cr
0.048& 2.16372& 96 &0.00157769\cr
\noalign{\hrule}
\noalign{\hrule}}}}
\par
\vbox{\offinterlineskip
In this table we list the angular velocity acquired by the DM
shell up to the point where the DM shells turns around.
This value is calculated
at the epoch of the turn around of each shell. The scale factor
at that turn around time is given in column 3. These quantities
are given for a range of initial texture collapse and initial
co-moving radius.}
\vfill \eject
\vbox{\moveright 0.5 in
\vbox{\offinterlineskip
\halign{\strut\vrule\quad # \quad & \vrule \hfil\quad # \hfil
& \quad \vrule \quad # \quad \vrule & \hfil\quad # \quad \hfil \vrule \cr
\noalign{\hrule}
Table 1 & & & \cr
\noalign{\hrule}
$a_i$ & $x_0$ & a/$a_{eq}$ & w ($ls^{-1}$) \cr
\noalign{\hrule}
0.064& 0.28702& 7.80306& 0.129639\cr
0.064& 1.4351& 65.0556& 0.00479661\cr
0.064& 2.8702& 128 &0.000734812\cr
\noalign{\hrule}
0.096& 0.426246& 14.3865 &0.0423705\cr
0.096& 2.13123 &119.955 &0.00165626\cr
0.096& 4.26246 &192 &0.000283749\cr
\noalign{\hrule}
0.128 &0.562777& 22.5143 &0.0190206\cr
0.128& 2.81388 &187.74 &0.00081579\cr
0.128& 5.62777 &256 &0.00015816\cr
\noalign{\hrule}
0.172& 0.746374 &36.1093 &0.00949166\cr
0.172& 3.73187& 301.128 &0.000447707\cr
0.172& 7.46374& 344 &9.5687e-05\cr
\noalign{\hrule}
0.256& 1.08455& 69.5543 &0.00365715\cr
0.256& 5.42276& 512 &0.000228605\cr
0.256& 10.8455& 512 &5.06313e-05\cr
\noalign{\hrule}
0.384& 1.57217 &138.656& 0.0014671\cr
0.384& 7.86085 &768 &9.83743e-05\cr
0.384& 715.7217& 768& 2.08143e-05\cr
\noalign{\hrule}
0.512& 2.03084& 228.674 &0.000952099\cr
0.512& 10.1542& 1024 &5.26307e-05\cr
0.512& 20.3084& 1024 &1.04759e-05\cr
\noalign{\hrule}
0.768& 2.87671& 467.535& 0.000655441\cr
0.768& 14.3835& 1536 &2.13841e-05\cr
0.768& 28.7671& 1536 &3.3304e-06\cr
\noalign{\hrule}
1.024& 3.64583& 779.061 &0.000399587\cr
1.024& 18.2291& 2048 &1.0869e-05\cr
1.024& 36.4583& 2048 &3.00253e-07\cr
\noalign{\hrule}
1.536& 5.01273& 1598.52& 0.000185055 \cr
1.536& 25.0637& 3072 &3.60041e-06\cr
1.536& 50.1273& 3072 &0 \cr
\noalign{\hrule}
2.048& 6.21222& 2655.68& 0.000103789 \cr
2.048& 31.0611& 4096& 9.4301e-07\cr
2.048& 62.1222& 4096& 0\cr
\noalign{\hrule}
2.99& 8.12198& 5168.26& 4.47294e-05 \cr
2.99& 40.6099& 5980& 0 \cr
2.99& 81.2198& 5980& 0 \cr
\noalign{\hrule}}}}

\vfill \eject
\newsec{Appendix 1}
Here we will derive the equation for the Harrison mechanism.

\eqn\Dconvective{{D_j \over D\tau} = {d \over dt}+ \vec{v}_j \cdot
\vec{\nabla} ={ \partial \over {\partial}t } + \vec{V}_j\cdot\vec{\nabla}}
 is the standard convective derivative
where ${d \over dt}$ follows the expansion and
$V_j = \dot{R}\vec{x} + R\dot{x}$. $\dot{R}\vec{x}$ is the hubble flow
 with the other term being the the peculiar velocity.
In an expanding Universe it must be replaced by

\eqn\DRv{{1 \over R} {D \over D\tau}\bigl(R\vec{v}\bigr) =
 {\partial\vec{v} \over {\partial}t} + {\dot{R} \over R}\vec{v}
  + {1 \over R} \bigl(\vec{V}\cdot\vec{\nabla}\vec{v} \bigr)}

where we often replace ${\dot{R} \over R}$= H , which is the hubble expansion
rate.

The equation for the electron motion is
\eqn\Eharrison{{\rho_e \over R} {D \over D\tau} \bigl( R\vec{v}_e \bigr) =
-{e n_e \over c} \bigl( \vec{E} + {1 \over c} \vec{V}_e \times \vec{B}
- {\vec{j} \over \sigma}\bigr)
 + {4 \over 3} \rho_{\gamma} n_e c
\sigma_{T_e} (\vec{v}_{\gamma} - \vec{v}_{e})
- {\nabla p \over R} - {\rho_e \nabla\phi \over R } }

the ion equation is
\eqn\Iharrison{{\rho_e \over R} {D \over D\tau} \bigl( R\vec{v}_p \bigr) =
+{e n_p \over c} \bigl( \vec{E} + {1 \over c} \vec{V}_p \times \vec{B}
- {\vec{j} \over \sigma}\bigr)
 + {4 \over 3} \rho_{\gamma} n_e c \sigma_{T_p} (\vec{v}_{\gamma} -
\vec{v}_{e})
- {\nabla p \over R} - {\rho_p \nabla\phi \over R } }
and the photon equation is

\eqn\pharrison{(\rho_{\gamma}+{1\over c^2}p_{\gamma}){1 \over R}
 {D \over D\tau} \bigl(
R\vec{v}_{\gamma} \bigr) = -{1 \over c^2} \vec{V}_{\gamma}{d\over dt}p_{\gamma}
-(\rho_{\gamma}+{1\over c^2}p_{\gamma})\nabla\phi
 - {4 \over 3}\rho_{\gamma} n_e c \sigma_{T_e} (\vec{v}_{\gamma} - \vec{v}_{e})
-{4 \over 3}\rho_{\gamma}n_p c\sigma_{T_p}(\vec{v}_{\gamma}
-\vec{v}_p) }

\eqn\dummyeqn{\nolabels +{4 \over 3}
{\rho_{\gamma}c \over 5 n \sigma}\nabla^2\vec{v}_{\gamma}}
We drop the last term, because the conductivity is very large
at that epoch. If we use the continuity equation
\eqn\Continuity{\vec{\nabla}\cdot\vec{V} = {-1 \over \bigl( \rho_{\gamma}
+{1 \over c^2}p_{\gamma}\bigr)} {d\rho_{\gamma} \over dt}}
and $p_{\gamma}={1 \over 3}\rho_{\gamma}c^2$ and
$\vec{\nabla}\cdot\vec{V}_{\gamma} = 3H + \vec{\nabla}\cdot\vec{v}$

we get

\eqn\Pharrison{\rho_{\gamma} {D \over D\tau}\vec{v}_{\gamma} = -{ 1\over 3}
\rho_{\gamma}\vec{V}_{\gamma} (\vec{\nabla}\cdot\vec{v}_{\gamma})
-\rho_{\gamma}\vec{\nabla}\phi -c\sigma_{T_e}\rho_{\gamma}n_e(\vec{v}_{\gamma}
-\vec{v}_e) -c\sigma_{T_p}\rho_{\gamma}n_p(\vec{v}_{\gamma}
-\vec{v}_p)}


Note that we will be taking the curl of \Eharrison \Iharrison and \Pharrison
so that the pressure and gravitational term will go away. A short calculation
of  $\vec{\nabla} \times $\Iharrison yields

\eqn\CIharrison{{1 \over R^2} {D \over D\tau}( R^2 \vec{\zeta}) +
\vec{\zeta}(\vec{\nabla}\cdot\vec{v}) -(\vec{\zeta}\cdot\vec{\nabla})\vec{v}
= -{n_p e \over \rho_p c}{1 \over R^2} {D \over D\tau}\bigl( R^2\vec{B} \bigr)
+{c \over 4\pi\sigma_{cond}}\Biggl( {1\over \mu} \vec{\nabla}^2 \vec{B}
- {\epsilon \over c^2 }\ddot{\vec{B}} \Biggr)
 - {4 \over 3}{\rho_{\gamma} \over \rho_p} n_p c
\sigma_{T_p} (\vec{\zeta}_{\gamma} -
\vec{\zeta}_{p}) }

where to find the RHS we assumed that $\vec{B}\cdot\nabla\vec{V} = H\vec{B}$
which is only exactly true at $\theta = {\pi \over 2}$ .
The conductivity $\sigma$ being very high, we can drop the wave equation
term in $\vec{B}$ . With
\eqn\Ai{\alpha_p = {n_p e \over \rho_i c} = {e \over m_p c} }
we can finally write $\nabla\times$\Iharrison as
\eqn\ICfharrison{{1 \over R} {D \over D\tau}\bigl[ R^2 \bigl(
\vec{\zeta}_p + \alpha_p\vec{B} \bigr)\bigr] =
-{4 \over 3} {\rho_{\gamma} \over m_p} c \sigma_{T_e} (\vec{\zeta}_{\gamma}
-\vec{\zeta}_e )}
similarly $\nabla\times$\Eharrison yields
\eqn\ECfharrison{{1 \over R} {D \over D\tau}\bigl[ R^2 \bigl(
\vec{\zeta}_e - \alpha_e\vec{B} \bigr)\bigr] = {4 \over 3}
{\rho_{\gamma} c \sigma_{T_p} \over m_e}\bigl( \vec{\zeta}_{\gamma}
- \vec{\zeta}_e \bigr)}

Right away we notice that the electron dynamics
is much more affected by the photon drag due to the smaller
electron mass on the denominator of the RHS of \ECfharrison .
Also notice that in \ICfharrison and \ECfharrison $\sigma_{T_e} /
m_p = \sigma_{T_p} / m_p$.
Taking the curl of \Pharrison yields
\eqn\CPharrison{\rho_{\gamma}{1 \over R} {D \over Dt}\bigl(
R\vec{\zeta}_{\gamma}\bigr) = - n_e c \sigma_{T_e} \rho_{gamma} \bigl(
2\vec{\zeta}_{\gamma} - \vec{\zeta}_e -\vec{\zeta}_p \bigr)}
where we have assumed $\vec{\nabla}\cdot\vec{v}$ is small .

(\ECfharrison + \ICfharrison )  yields

\eqn\finalharrison{{1 \over R} {D \over D\tau}\bigl[ R^2 \bigl(
\vec{\zeta}_p+\vec{\zeta}_p \bigr)\bigr] =
{1 \over R} {D \over D\tau}\bigl[ R^2 \bigl(
\alpha_e\vec{B} \bigr)\bigr] }

\par
We have used $\alpha_e \gg \alpha_p$. Since the proton will
spin down faster than the electron, the proton
vorticity will be small. Integrating \finalharrison yields
the wanted result \BHarrison .

This
was derived under the assumption that the matter background is
spinning down faster from having the same initial angular velocity
as the radiation.
In the original Harrison mechanism, the vortices were primordial and the
radiation had the same initial velocity as the matter.
The evolution was different because for the radiation
(and the electrons tied to it by the strong Thompson cross section)
$ L = \rho V r^2 \omega \sim a^{-4} a^3 a^2 \omega = constant $ so
$\omega_{\gamma} \sim a^{-1}$ where a is the scale factor, whereas for the
matter $w \sim a^{-2}$ .
\par
In our model the angular velocity will be provided by angular
torquing (+ dissipation). We have to require that the angular
momentum gained by the matter has had time to spin down.
For angular velocity acquired by gravitational
torquing ( + dissipation) ,
most of the angular momentum of the matter should be acquired
by the time the dark matter shell has reached it's turn around
radius.
Before recombination the photon mean free path
 $l_{\gamma}={1 \over n_e \sigma_T}$ is $6 \times 10^{-5} Mpc
\Omega_0^{-1} h_{50}^{-1}
({1100 \over 1+z})^3 $ so that density perturbation of the radiation
by textures don't have time to diffuse out of their creation
region before recombination and they will
acquire angular velocity to due gravitational
torquing+dissipation just as
the matter did. So at a given shell radius,
the angular momentum growth must peak before
recombination ( $a \sim 5$ )
in order to have a sizable magnetic by the Harrison
mechanism.

\vfill \eject

\listrefs


\end